\documentclass[fonsize=10pt,a4paper, egregdoesnotlikesansseriftitles]{scrartcl}
\usepackage{tocloft}
\usepackage{amsmath,amssymb,amsfonts}
\usepackage{algorithmic}
\usepackage{graphicx}
\usepackage{textcomp}
\usepackage{url}
\usepackage{upgreek}
\DeclareSymbolFont{extraup}{U}{zavm}{m}{n}
\DeclareMathSymbol{\vardiamondsuit}{\mathalpha}{extraup}{87}

\usepackage{longtable,booktabs,array}
\usepackage{calc} 

\usepackage{amsmath,amssymb}
\usepackage{lmodern}
\usepackage{iftex}
\usepackage{authblk}

\makeatother
\usepackage{xcolor}
\usepackage{graphicx}
\makeatletter

\providecommand{\tightlist}{%
  \setlength{\itemsep}{0pt}\setlength{\parskip}{0pt}}
\IfFileExists{bookmark.sty}{\usepackage{bookmark}}{\usepackage{hyperref}}
\IfFileExists{xurl.sty}{\usepackage{xurl}}{} 
\hypersetup{
  hidelinks,
  pdfcreator={LaTeX via pandoc}}

\usepackage[margin=1.2in]{geometry}

\setkeys{Gin}{width=0.8\linewidth,height=\textheight,keepaspectratio}
\makeatletter

\usepackage[backend=biber,url=false,eprint=false,style=ieee,maxbibnames=3]{biblatex}

\cftpagenumbersoff{figure}
\cftpagenumbersoff{table}     

\begin{document}

\title{\Large Comparison of Consecutive and Re-stained Sections for Image Registration in Histopathology}
\author[a,*]{\small Johannes Lotz}
\author[a,*]{\small Nick Weiss}
\author[b,c]{\small Jeroen van der Laak}
\author[a]{\small Stefan Heldmann}
\affil[a]{Fraunhofer Institute for Digital Medicine MEVIS, Lübeck, Germany}
\affil[b]{Department of Pathology, Radboud University Medical Center, Nijmegen, The Netherlands}
\affil[c]{Center for Medical Image Science and Visualization, Linköping University, Linköping, Sweden}

\date{\noindent \footnotesize\textbf{*}JL and NW contributed equally, correspondence to johannes.lotz@mevis.fraunhofer.de }

\setcounter{Maxaffil}{0}
\renewcommand\Affilfont{\itshape\small}


\maketitle
\setcounter{page}{1}

\begin{abstract}

Purpose: In digital histopathology, virtual multi-staining is important for diagnosis and biomarker research. Additionally, it provides accurate ground-truth for various deep-learning tasks. Virtual multi-staining can be obtained using different stains for consecutive sections or by re-staining the same section. Both approaches require image registration to compensate tissue deformations, but little attention has been devoted to comparing their accuracy. 

Approach: We compare variational image registration of consecutive and re-stained sections and analyze the effect of the image resolution which influences accuracy and required computational resources. We present a new hybrid dataset of re-stained and consecutive sections (HyReCo, 81 slide pairs, approx. 3000 landmarks) that we made publicly available and compare its image registration results to the automatic non-rigid histological image registration (ANHIR) challenge data (230 consecutive slide pairs). 

Results: We obtain a median landmark error after registration of 7.1 $\upmu$m (HyReCo) and 16.0 $\upmu$m (ANHIR) between consecutive sections. Between re-stained sections, the median registration error is 2.3 $\upmu$m and 0.9 $\upmu$m in the two subsets of the HyReCo dataset. We observe that deformable registration leads to lower landmark errors than affine registration in both cases, though the effect is smaller in re-stained sections. \\
Conclusion: Deformable registration of consecutive and re-stained sections is a valuable tool for the joint analysis of different stains. 

Significance: While the registration of re-stained sections allows nucleus-level alignment which allows for a direct analysis of interacting biomarkers, consecutive sections only allow the transfer of region-level annotations. The latter can be achieved at low computational cost using coarser image resolutions. 

\end{abstract}






\hypertarget{introduction}{%
\section{Introduction}\label{introduction}}

In histopathology, much insight into disease subtyping, biomarker
discovery, and tissue organization is gained by analyzing differently
stained histological sections. For this procedure, a fixed tissue is
transferred into a paraffin block and cut into 2--5 \textmu m thin
slices. These slices are subsequently stained by
e.g.~immunohistochemistry, and---in a digital workflow---scanned to
obtain a digital whole slide image (WSI)
\autocite{mukhopadhyay_whole_2017}. The resulting image can be used for
digital analysis, e.g.~in biomarker discovery by combining two or more
different stains~\autocite{harder_automatic_2019}. Deep-learning models
are increasingly used to analyze histopathology slides and first methods
have been cleared for clinical use \autocite{van_der_laak_deep_2021}.
These methods require a large amount of annotated images to learn
specific tissue properties. Image registration is used to automatically
create annotations as training data in order to reduce the time spent on
manually annotating slide
images~\autocite{bulten_epithelium_2019,tellez_whole-slide_2018}.

Enabled by digital slide scanners, a re-staining approach which was
initially used in fluorescence microscopy and known as tissue-based
cyclic immunofluorescence (t-CyCIF)
\autocite{gerdes_highly_2013,lin_highly_2018} is gaining popularity in
bright-field imaging
\autocite{remark_-depth_2016,bulten_epithelium_2019,banik_high-dimensional_2020}.
Instead of staining consecutive sections and scanning them later, a
section is stained and scanned first. In a second step, the stain is
washed or bleached and another stain is applied. After re-scanning, both
images contain the same tissue with different staining, so that it is
possible to compare the same cell with respect to different antibodies
or markers. However, we still observe nonlinear deformations in the
tissue, which are most likely due to the chemical reactions during the
re-staining process.

Independently of the sectioning method, researchers face a number of
questions when applying image registration to a new dataset. These
include

\begin{itemize}
\tightlist
\item
  Which image resolution is best suited to obtain the best accuracy
  while keeping the computational cost as low as possible?
\item
  Is deformable registration required or is an affine registration
  sufficient, especially for the registration of re-stained sections
  where little differences are expected between both images?
\end{itemize}

As we show in this work, image registration is required in both,
consecutive as well as re-stained, image pairs. We further compare the
accuracy of the registration for both types of image pairs which---to
our knowledge---has not been analyzed before. In the case of
registration of re-stained sections, accuracy is achieved at the nucleus
level. In the case of consecutive sections, this level of accuracy
cannot usually be reached due to the lack of corresponding objects at
the appropriate resolution caused by the slice thickness or distance.
Here, a good registration of structures with a size above the nucleus
level can be achieved on the basis of images with relatively low
resolution and the use of a nonlinear deformation model.

We compare the two types of image pairs using an optimization-based
image registration method that is based on minimizing an energy
functional consisting of a distance measure and a regularizer
\autocite{modersitzki_fair_2009}. This class of optimization-based
methods is widely used in medical imaging
\autocite{keszei_survey_2017,sotiras_deformable_2013} and has also been
applied to problems in pathology
\autocite{lotz_patch-based_2016,pichat_survey_2018,borovec_anhir_2020,wodzinski_multistep_2020,venet_accurate_2021}.

This class of energy-minimizing methods makes explicit model assumptions
through the choice of distance measure and regularization scheme. When
applying a method to a new dataset, model refinements can be made by
adjusting its parameters. For example, when a new dataset contains
larger deformations, the weight that balances image distance and
regularization can be adapted to allow for larger displacements.

Another class of methods that gained popularity in the recent years is
based on training a deep learning model to estimate the deformation in
problems in medical imaging \autocite{haskins_deep_2020} and
specifically pathology \autocite{wodzinski_deephistreg_2021}. Here, the
model assumptions are made implicitly by the training data. This in turn
makes generalization and adaption to unseen datasets more challenging,
although recent work
\autocite{jiang_multi-scale_2020,fu_lungregnet_2020,hering_cnn-based_2021}
addresses this issue.

Below, we first describe the used registration method and its
application to re-stained and consecutive slide images. We then describe
an evaluation framework based on landmarks accuracies on two datasets,
the ``automatic nonlinear histological image registration challenge''
(ANHIR\footnote{https://anhir.grand-challenge.org/},
\autocite{borovec_anhir_2020}) and on a new dataset ``HyReCo''
\autocite{van_der_laak_hyreco_2021} that contains both consecutive and
re-stained slides and that we make publicly available. Finally, we
analyze the accuracy of the image registration method with respect to
image resolution and sectioning in both datasets.

\hypertarget{fully-automatic-image-registration}{%
\section{Fully-automatic image
registration}\label{fully-automatic-image-registration}}

We compare the registration of the two sectioning methods based on a
3-step, energy-minimizing registration pipeline. It consists of 1) a
robust pre-alignment, 2) an affine registration computed on coarse
resolution images, and 3) a curvature-regularized deformable
registration. The method is based on the variational image registration
framework first described by Fischer and Modersitzki
\autocite{fischer_fast_2003,modersitzki_fair_2009} which has been
applied to many clinical fields from histology
\autocite{schmitt_image_2007} to radiology
\autocite{ruhaak_highly_2013,konig_matrix-free_2018-1}.

Given a so-called reference image \(R : {\mathbb{R}}^2\to{\mathbb{R}}\)
and a so-called template image \(T: \mathbb{R}^2\to\mathbb{R}\), the
goal of image registration is to find a reasonable spatial
transformation \(y:{\mathbb{R}}^2\to{\mathbb{R}}^2\) such that
\(R({\mathbf{x}}) \approx T(y({\mathbf{x}}))\), i.e., \(R\) and the
deformed template \(T\circ y\) are similar in an adequate sense.

Following \autocite{modersitzki_fair_2009}, we formulate the image
registration as the optimization problem
\(J(R, T, y) \stackrel{y}{\to}\min\) of an appropriate objective
function \(J\) with respect to the desired spatial transformation. A key
component of the objective function is a so-called distance or image
similarity measure that quantifies the quality of the alignment. We use
the Normalized Gradient Fields (NGF) distance measure
\autocite{haber_intensity_2007} as it has been shown to be robust to
different stains and is suitable for multi-modal image registration of
histological images \autocite{bulten_epithelium_2019}. For the
discretization, 2D images with extents \(n_1\)-by-\(n_2\) are assumed,
correspondingly consisting of \(N=n\_1\cdot n\_2\) pixels with uniform
size \(h \in \mathbb R\) in each dimension and pixel centers
\({\mathbf{x}}_1,...,{\mathbf{x}}_N;\, \mathbf x\_i \in \mathbb R^2\).
The NGF distance measure is given by

\[
\begin{aligned}
    \text{NGF}(R,T,y) &=  \\\\
    \quad \frac{h^2}{2} \cdot \sum_{i=1}^N 1 &- \left( \frac{\langle\nabla T(y({\mathbf{x}}_i)), \nabla R({\mathbf{x}}_i) \rangle_\varepsilon  }{\| \nabla T(y({\mathbf{x}}_i))\|_\varepsilon  \, \| \nabla R({\mathbf{x}}_i)\|_\varepsilon} \right)^2
\end{aligned}
\]

with
\(\langle{\mathbf{x}},{\mathbf{y}}\rangle_\varepsilon={\mathbf{x}}^\top{\mathbf{y}}+\varepsilon^2\),
\(\|{\mathbf{x}}\|_\varepsilon := \sqrt{\langle {\mathbf{x}},{\mathbf{x}}\rangle_\varepsilon}\),
and the edge parameter \(\varepsilon\), which controls the sensitivity
to edges in contrast to noise. This image distance becomes minimal if
intensity gradients and edges, respectively, are aligned and which
therefore leads to the alignment of morphological structures.

The NGF distance measure is used in all three steps of the registration
pipeline: Pre-alignment, affine registration, and deformable
(non-linear) registration. In addition, we use a multilevel optimization
scheme that starts with the registration of images at low resolution
levels and then refines the transformation to higher image resolutions
to reduce the risk of converging too early to local minima and to speed
up the optimization process \autocite{haber_multilevel_2006}. The
per-level optimization is performed using a Gauss-Newton type (affine
registration) and L-BFGS quasi-Newton (deformable registration) method,
see e.g. \autocite{modersitzki_fair_2009} or
\autocite{song_review_2017,nocedal_numerical_2006} for a more detailed
discussion and additional strategies.

\begin{longtable}[]{@{}
  >{\raggedright\arraybackslash}p{(\columnwidth - 2\tabcolsep) * \real{0.6076}}
  >{\raggedright\arraybackslash}p{(\columnwidth - 2\tabcolsep) * \real{0.3924}}@{}}
\caption{Parameters used in the registration pipeline for all datasets.
\label{tbl:parameters}}\tabularnewline
\toprule()
\begin{minipage}[b]{\linewidth}\raggedright
Registration step
\end{minipage} & \begin{minipage}[b]{\linewidth}\raggedright
Parameter values
\end{minipage} \\
\midrule()
\endfirsthead
\toprule()
\begin{minipage}[b]{\linewidth}\raggedright
Registration step
\end{minipage} & \begin{minipage}[b]{\linewidth}\raggedright
Parameter values
\end{minipage} \\
\midrule()
\endhead
\textbf{Step 1: Pre-Alignment} ~~~~~~~~ & \\
No.~of levels \(N_{\text{level}}\) & 4 \\
No.~of rotations \(N_{\text{rot}}\) & 32 \\
image resolution (\(\frac{\upmu m}{\text{px}}\)) & approx. 200 \\
image size (px & approx.~100×200 \\
NGF \(\varepsilon\) & 0.1 \\
\textbf{Step 2: Affine} & \\
image resolution (\(\frac{\upmu m}{\text{px}}\)) & approx. 248 -- 1 \\
image size (px) & approx.~100×200 -- 25k×55k \\
No.~of levels \(N_{\text{level}}\) & 3 (248
\(\frac{\upmu m}{\text{px}}\)) -- 11 (1
\(\frac{\upmu m}{\text{px}}\)) \\
NGF \(\varepsilon\) & 0.1 \\
\textbf{Step 3:~Deformable} & \\
image resolution (\(\frac{\upmu m}{\text{px}}\)) & approx. 248 -- 1 \\
image size (px) & approx.~100×200 -- 25k×55k \\
No.~of levels \(N_{\text{level}}\) &
3~(248~\(\frac{\upmu m}{\text{px}}\))-- 11~(1
\(\frac{\upmu m}{\text{px}}\)) \\
NGF \(\varepsilon\) & 0.1 \\
regularizer weight \(\alpha\) & 0.1 \\
control point grid \(m\) & 257×257 nodes \\
\bottomrule()
\end{longtable}

All of the three following registration steps rely on the edge parameter
\(\varepsilon\), the number of levels \(N_{\text{level}}\) of the image
pyramid, and the image resolution at the finest level. The parameters
are set independently for each step and such that the registration error
is minimal and the deformation grid is regular in the sense that it is
not folded in the image domain. These parameters are shown in
Table~\ref{tbl:parameters}.

\hypertarget{step-1-automatic-rotation-alignment-ara}{%
\subsection{Step 1: Automatic rotation alignment
(ARA)}\label{step-1-automatic-rotation-alignment-ara}}

Before histological images are scanned, the tissue is cut, preprocessed,
and stained in a pathology lab.

After this manual process, neighboring tissue slices can end up in
arbitrary positions on the object slide (such as upside down or turned
in various ways). In general, no assumptions can be made on the initial
tissue positioning and---in a first step---we aim to find a rigid
alignment, correcting for global translation and global rotation.

Images are assumed to be available in a multilevel image data format to
reduce the time and memory requirements to load the image data at a
given resolution.

The NGF distance measure is based on structural changes expressed
through the image gradient and therefore, color information is of
limited value. To reduce the amount of image data to be handled, all
images are converted from color to gray scale and inverted to obtain a
black background while loading from disk.

Automatic Rotation Alignment (ARA) first determines the center of mass
\autocite{beatty_principles_1986} of both images, using the gray values
of the pixels as the weights. Let \((t_{1}, t_{2})\) be the vector
pointing from the center of mass of the reference image to the center of
mass of the template image, and let
\(\phi_k=2\pi (k-1)/(N_\text{rotations}-1)\),
\(k=1,\hdots,N_\text{rotations}\) be equidistant rotation angles
sampling the interval \([0,2\pi)\). For each angle, a rigid registration
is computed, optimizing \[
\begin{aligned}
J(R, T, y_{\text{rigid}}) &= \text{NGF}(R,T, y_{\text{rigid}})\to\min, \\\\
y_{\text{rigid}} : \mathbb R^2 &\mapsto \mathbb R^2, y_{\text{rigid}} \text{ parameterized by } (\phi_k,t_1,t_2)
\end{aligned}
\] with initial parameters \((\phi_k,t_1,t_2)\),
\(k=1,\hdots,N_\text{rotations}\). Among all \(N_\text{rotations}\)
rigid registration results, the minimizer \(y_{\text{rigid}}^*\) with
the smallest image distance is selected as an initial guess for the
subsequent affine registration.

\hypertarget{step-2-affine-registration}{%
\subsection{Step 2: Affine
registration}\label{step-2-affine-registration}}

In a second step, again an NGF-based image registration is computed. To
allow for additional degrees of freedom, the registration is optimized
with respect to an affine transformation \(y_{\text{affine}}\) and based
on a finer image resolution with the previously computed
\(y_{\text{rigid}}^*\) as an initial guess. The resulting transformation
is then used as initial guess for a subsequent deformable registration.

\hypertarget{step-3-deformable-registration}{%
\subsection{Step 3: Deformable
registration}\label{step-3-deformable-registration}}

The final step is a deformable image registration. Here, the
transformation \(y\) is given by \[
y({\mathbf{x}}) = {\mathbf{x}}+ u({\mathbf{x}})
\] with so-called displacement \(u:{\mathbb{R}}^2\to{\mathbb{R}}^2\),
\(u = (u_1, u_2)\) \autocite{modersitzki_fair_2009}.

In contrast to an affine registration, the deformation is not restricted
to a particular parameterizable deformation model and the nonlinear
transformation is controlled by introducing a regularization term into
the objective function that measures the deformation energy and
penalizes unwanted transformations. Here we use the so-called curvature
regularization, which penalizes second-order derivatives of the
displacement \autocite{fischer_curvature_2003} and which has been shown
to work very well in combination with the NGF distance measure
\autocite{ruhaak_highly_2013,konig_matrix-free_2018-1}. As with the NGF
distance, we evaluate the displacements in the pixel centers
\({\mathbf{x}}_1,\hdots,{\mathbf{x}}_m\) with uniform grid spacing \(h\)
and use finite differences to approximate the derivatives. Thus, the
discretized curvature regularizer is defined as \[
\text{CURV}(y) = \frac{h^2}2 \sum_{i=1}^{m} |\Delta^h u_1({\mathbf{x}}_i)|^2 + |\Delta^h u_2({\mathbf{x}}_i)|^2
\]

where \(\Delta^h\) is the common 5-point finite difference approximation
of the 2D Laplacian \(\Delta = \partial_{xx}+\partial_{yy}\) with
Neumann boundary conditions. In summary, for deformable registration, we
minimize the objective function \[
J(R, T, y) := \text{NGF}(R,T, y) + \alpha \text{CURV}(y)\to\min,
\] with respect to the deformation \(y\) with the previously computed
optimal \(y_{\text{affine}}\) as initial guess. The parameter
\(\alpha>0\), is a regularization parameter that controls the smoothness
of the computed deformation. The parameter \(\alpha\) is chosen manually
to achieve a smooth deformation and avoid topological changes (lattice
folds), while being flexible enough to correct for local changes that
improve image similarity. The resolution of the control point grid is
independent of the image resolution and is typically chosen to be
coarser than the image resolution (see also Table \ref{tbl:parameters}).
A higher number of grid points allows for a more accurate representation
of local deformations. Linear interpolation is used to evaluate the
deformation between its grid nodes.

\begin{figure}
\centering
\includegraphics{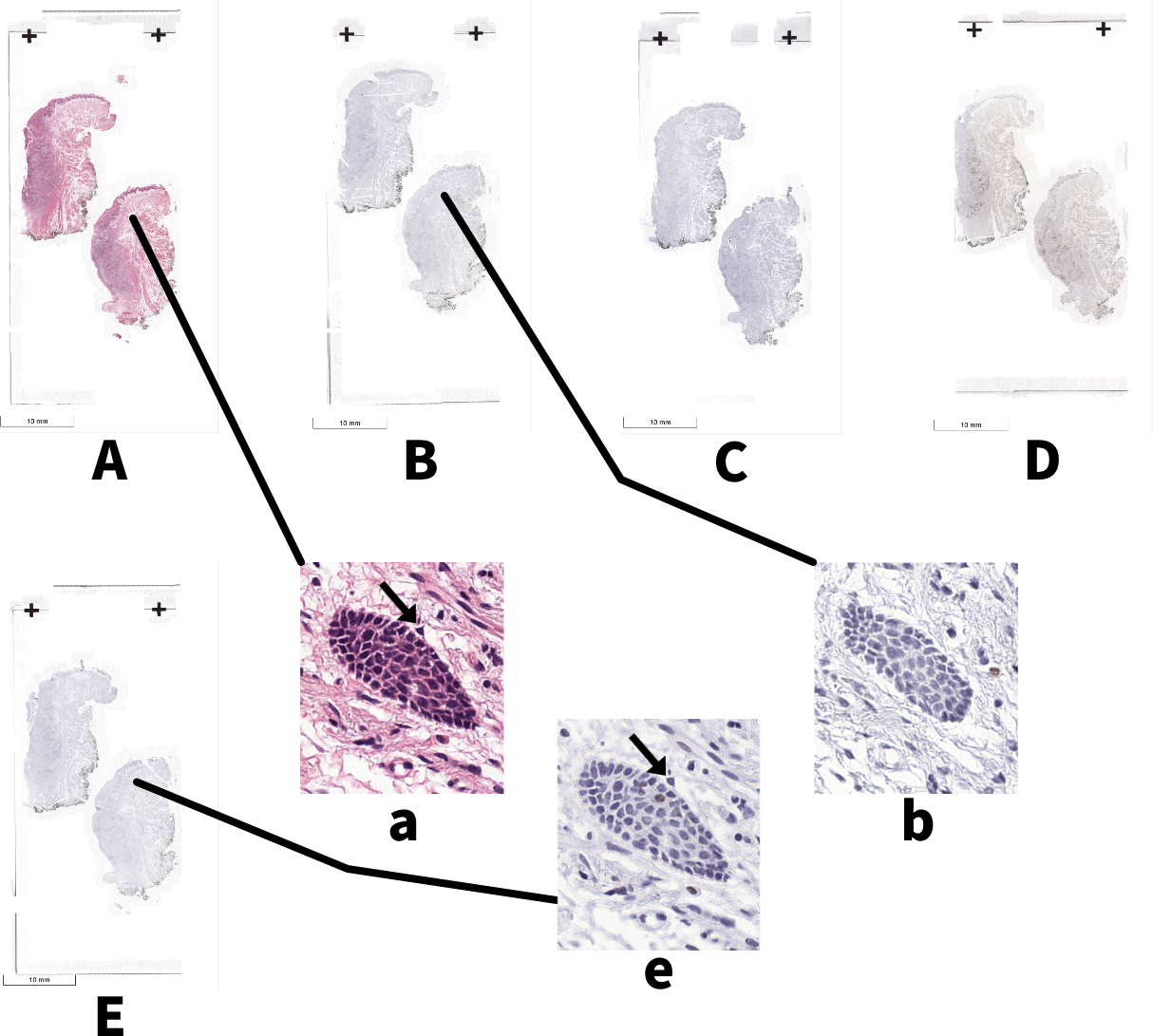}
\caption{One of nine sets of the HyReCo data. Slides A--D are
consecutive stains (H\&E, CD8, Ki67, CD45RO). Slide E is washed and
re-stained from slide A. In the cropped images a, b, and e corresponding
nuclei can be found only between the re-stained slide pair a and
e.\label{fig:hyreco}}
\end{figure}

\begin{figure}
\centering
\includegraphics{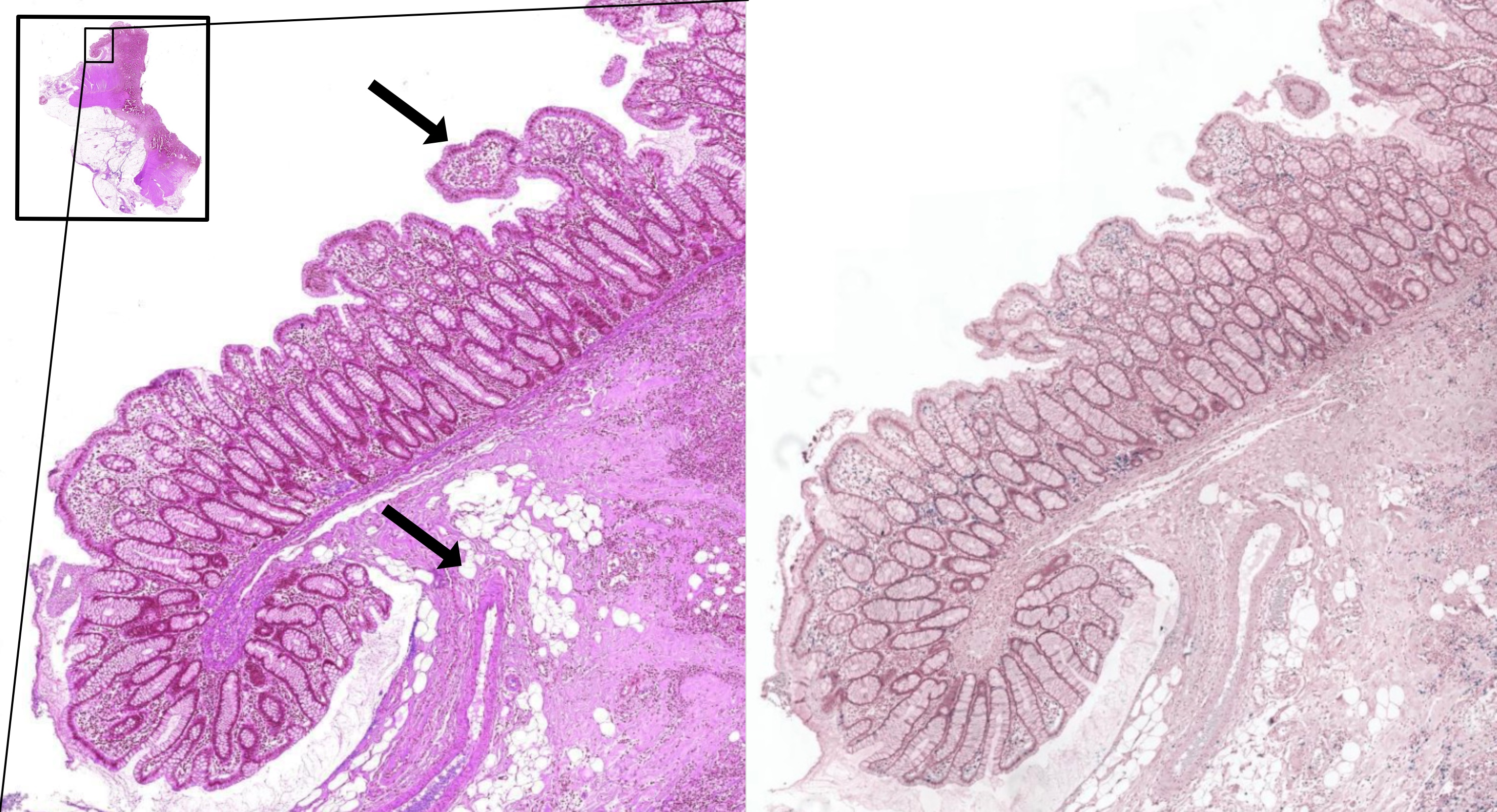}
\caption{Two consecutive slides from the ANHIR dataset (image set COAD
03). Structures that are only present in one image and that cannot be
aligned by image registration are indicated by
arrows.\label{fig:anhir_structure}}
\end{figure}

\hypertarget{evaluation}{%
\section{Evaluation}\label{evaluation}}

We compare both image acquisition methods with respect to the accuracy
of the registration in a new, previously unpublished dataset,
HyReCo~\autocite{van_der_laak_hyreco_2021}, that combines re-stained and
consecutive sections. To relate to the previous work in registration of
consecutive sections, we addtionally evaluate the registration accuracy
in the training part of the ANHIR challenge
data~\autocite{borovec_benchmarking_2018-1}.

We report the distribution of the target registration error
\(\| \mathbf r_k - \mathbf t_k \|_2, \, k=1,...,N_\text{images}\) and
its median \[
\operatorname{MTRE} = \text{median}_k \left(\| \mathbf r_k - \mathbf t_k \|_2 \right)
\] over all \(N_\text{images}\) image pairs and over all available
landmarks \(\mathbf r_k, \mathbf t_k \in \mathbb R^2\) in both datasets.

Multiple parameterizations were tested systematically and the parameter
set with the lowest \({\operatorname{MTRE}}\) was selected (Table
\ref{tbl:parameters}). Moderate modifications in NGF \(\varepsilon\),
regularizer parameter \(\alpha\), size of the deformation grid and
number of levels only show a small influence on the accuracy when
registering coarse image resolutions (up to approx. 4 \(\upmu\)m/px). On
higher image resolutions, the parameter choice seems to have a larger
impact. We choose the parameters reported in Table \ref{tbl:parameters}
that lead to the best results across all datasets. While this
parameterization was optimal in the median across all images, single
registrations can be improved by determining an individual set of
parameters.

The registration was applied to consecutive and re-stained sections in
the HyReCo and ANHIR datasets.

\hypertarget{hybrid-re-stained-and-consecutive-data-hyreco}{%
\subsubsection{Hybrid Re-stained and Consecutive Data
(HyReCo)}\label{hybrid-re-stained-and-consecutive-data-hyreco}}

The HyReCo dataset was acquired at the Radboud University Medical
Center, Nijmegen, the Netherlands\footnote{The requirement for ethical
  approval was waived by the IRB of Radboudumc, Nijmegen, the
  Netherlands, under file number 2020-6972.}.

\hypertarget{hyreco-subset-a-re-stained-consecutive}{%
\paragraph{HyReCo subset A (re-stained \&
consecutive)}\label{hyreco-subset-a-re-stained-consecutive}}

It consists of two subsets of slides, first (A) nine sets of consecutive
sections, each containing four slides stained with H\&E, CD8, CD45RO,
Ki67, respectively (Fig. \ref{fig:hyreco}). In addition, PHH3-stained
slides have been produced by removing the cover slip from the respective
H\&E-stained slide, bleaching the H\&E stain, re-staining the same
section with PHH3 and scanning it again, similar to the t-CyCIF
technique \autocite{gerdes_highly_2013,lin_highly_2018} that is well
established in fluorescence imaging. For each of these sections, 11--19
landmarks (138 per stain, 690 in total) have been placed manually on
corresponding structures and verified by two experienced researchers.

Finding the same points across several consecutive slides is quite
difficult, because care must be taken to locate a similar point in all
slides of the stack simultaneously. In contrast to these consecutive
sections, an image pair of re-stained sections contains the same cells
and nuclei such that a one-to-one correspondence can be found for most
structures.

\hypertarget{hyreco-subset-b-re-stained}{%
\paragraph{HyReCo subset B
(re-stained)}\label{hyreco-subset-b-re-stained}}

To overcome the limitations in annotation accuracy imposed by the
simultaneous annotation of consecutive and re-stained slides, a second
subset (B) of re-stained slides without corresponding consecutive
sections were scanned and annotated. An additional number of 2303
annotations were produced for 54 additional image pairs of H\&E-PHH3
(approx. 43 annotations per pair). These have again been verified by two
experienced researchers.

All images have been digitized with a resolution of 0.24~\(\upmu\)m/px
and are approximately \(95000 \times 220000\) pixels in size at their
highest magnification level.

To estimate a lower bar for landmark accuracy, two researchers annotated
the same structures (approx. 20 landmarks each) in the same and in one
consecutive slide, independently from each other. In this setting, the
inter-observer error on the same section was 0.57~\(\upmu\)m ± 0.36
(mean ± standard deviation), corresponding to 2.3 pixels ± 1.5 and the
intra-observer error was in a similar range (0.53~\(\upmu\)m ± 0.32). In
two consecutive sections (H\&E and Ki67) the inter-observer error was
1.1~\(\upmu\)m ± 0.6 (4.7~pixels ± 2.6). In the consecutive sections,
the landmark positions were selected such that a corresponding structure
was available in both images. For many structures this is not always the
case in consecutive sections such that the inter-observer error likely
overestimates the possible alignment accuracy.

The dataset including the landmarks has been made available at
{[}\textcite{van_der_laak_hyreco_2021}{]}\footnote{https://dx.doi.org/10.21227/pzj5-bs61}
under the Creative Commons Attribution-ShareAlike 4.0 International
license\footnote{https://creativecommons.org/licenses/by-sa/4.0/}.

\hypertarget{anhir-dataset}{%
\subsubsection{ANHIR Dataset}\label{anhir-dataset}}

The accuracy of the registration of serial sections depends on the
distance between the sections and on the quality of the tissue
sectioning. To broaden the scope of the analysis and to make the results
comparable to previous work in registration of serial sections, we
additionally evaluate the accuracy of the registration of the ANHIR
challenge data~\autocite{borovec_benchmarking_2018-1}.

The public part of the ANHIR challenge dataset consists of 230 image
pairs from 8 different tissue types (lung lesions, whole mice lung
lobes, mammary glands, mice kidney, colon adenocarcinoma, gastric mucosa
and adenocarcinoma, human breast, human kidney) with 18 different
stains. An example is shown in Fig. \ref{fig:anhir_structure}.

In the following sections, we measure the accuracy of deformable and
affine registration with respect to image resolution on both datasets.
We distinguish re-stained and consecutive sectioning and determine the
possible alignment accuracies in the different datasets.

\hypertarget{results}{%
\section{Results}\label{results}}

We apply the 3-step registration pipeline to the HyReCo datasets and to
the ANHIR training dataset.

\hypertarget{experiment-1-images-resolution}{%
\subsection{Experiment 1: Image's
Resolution}\label{experiment-1-images-resolution}}

\begin{figure}
\centering
\includegraphics{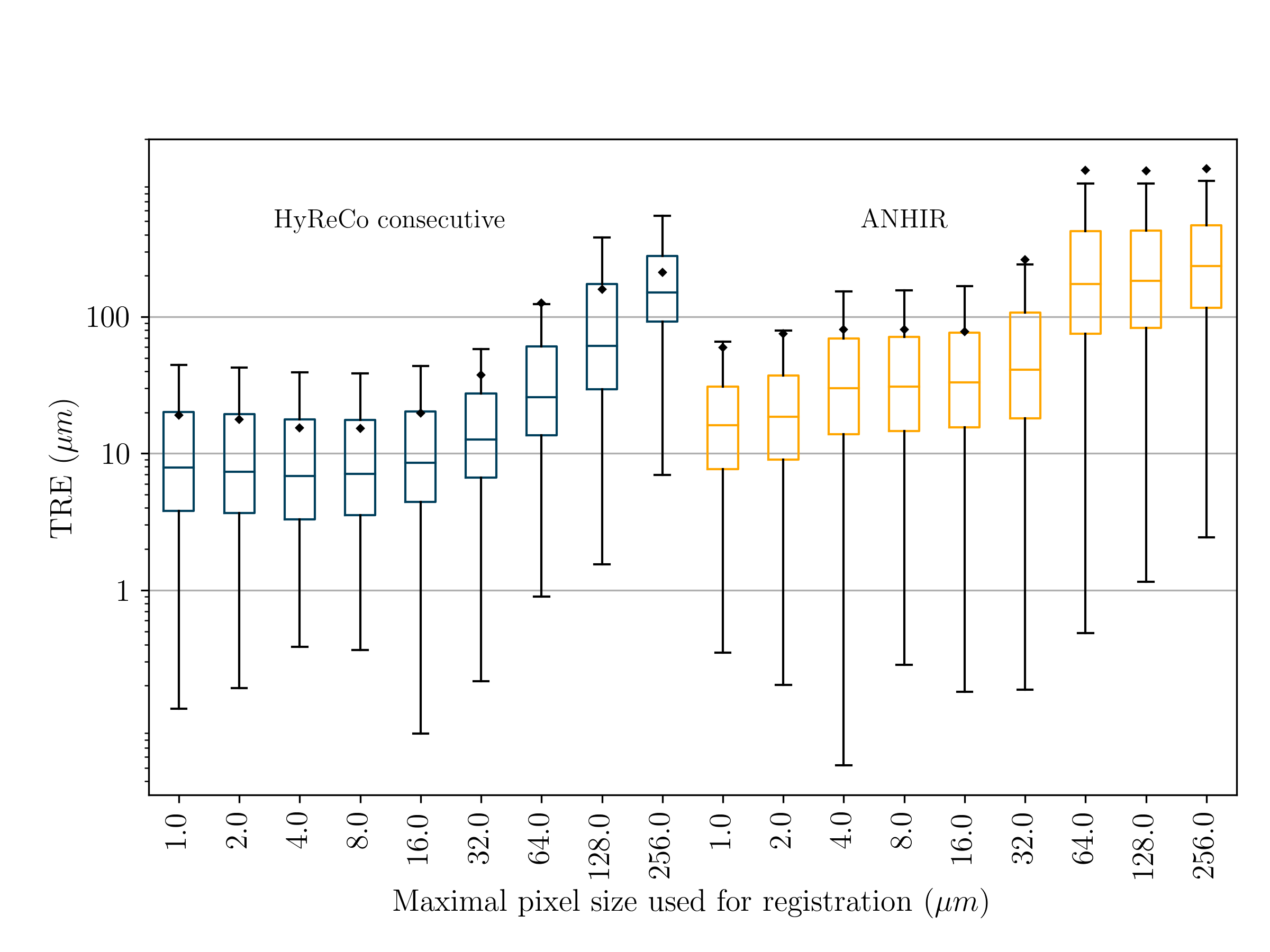}
\caption{TRE after \textbf{consecutive} deformable registration at
different image resolutions for the HyReCo (left group, blue) and the
ANHIR dataset (right group, yellow, logarithmic plot). The HyReCo
dataset shows overall smaller registration errors. The boxes denote the
interquartile range and the whiskers extend this range by a factor of
1.5. The diamond (\(\vardiamondsuit\)) denotes the mean TRE.
\label{fig:resultsdeformableconsecutive}}
\end{figure}

\begin{figure}
\centering
\includegraphics{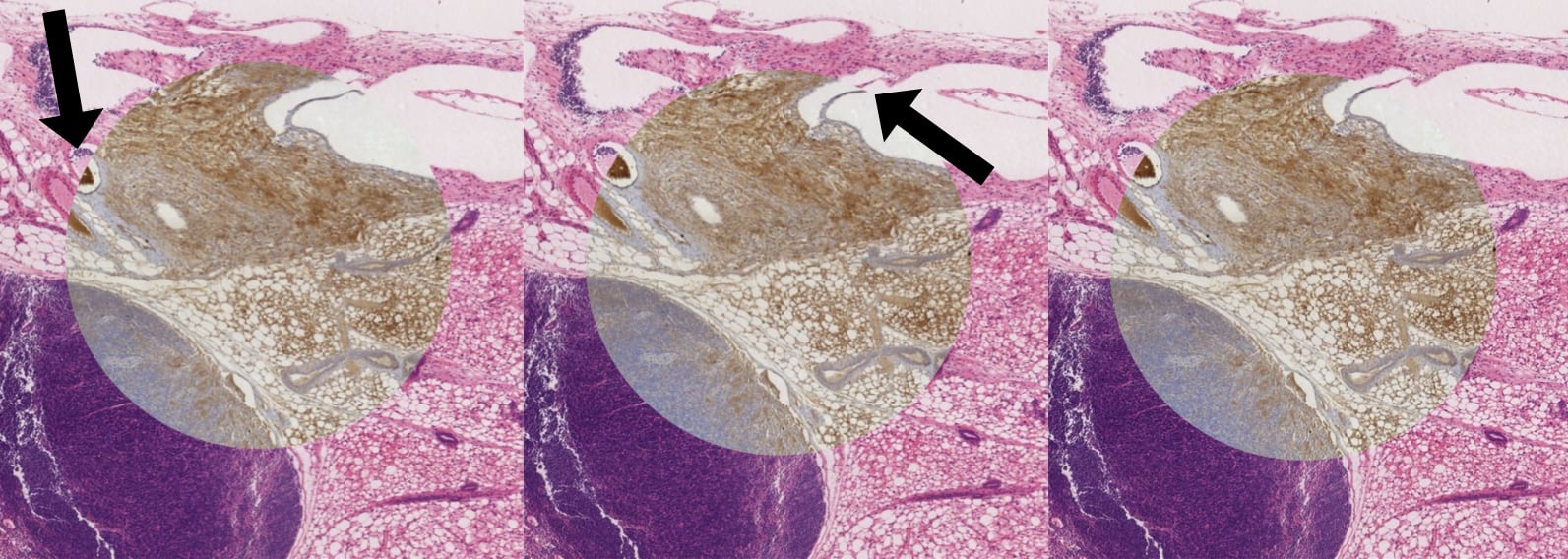}
\caption{Spy-view of an image pair from the ANHIR dataset after
pre-alignment, affine and deformable registration (left to right).
Arrows indicate tissue regions with misalignment.
\label{fig:images_anhir}}
\end{figure}

\begin{figure}
\centering
\includegraphics{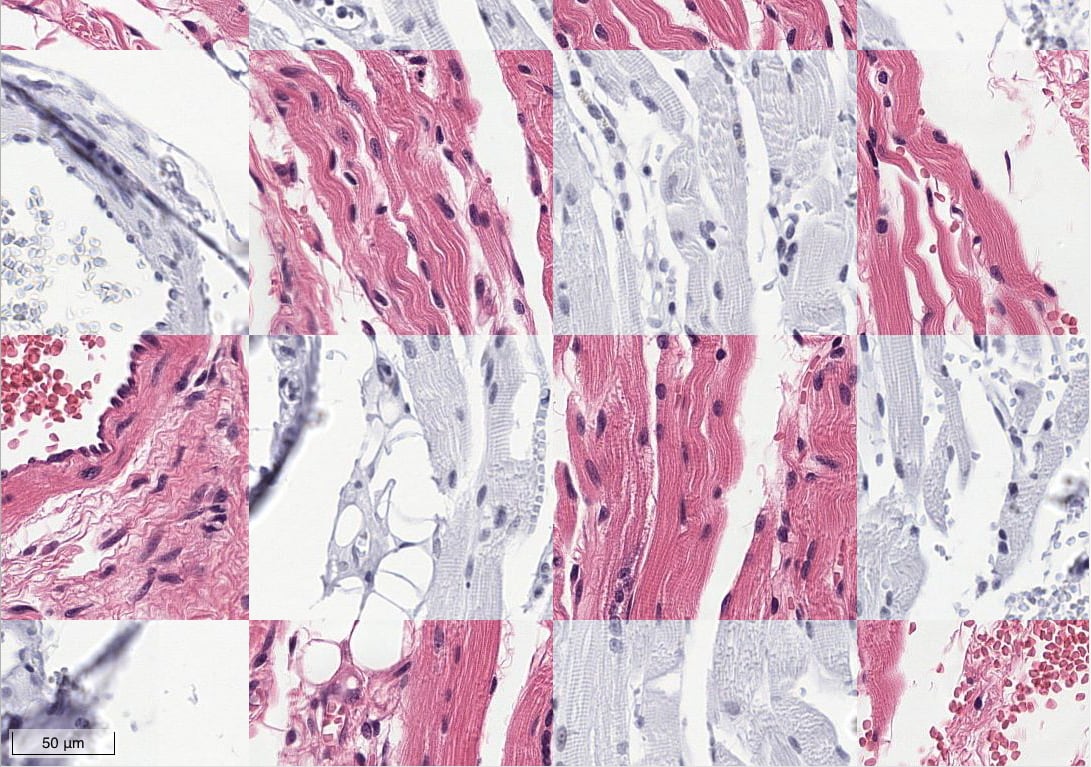}
\caption{Checkerboard plot after registration of a re-stained image
pair. Nucleus correspondences are visible at the borders of the
checkerboard tiles. \label{fig:checkerboard_re-stained}}
\end{figure}

\begin{figure}
\centering
\includegraphics{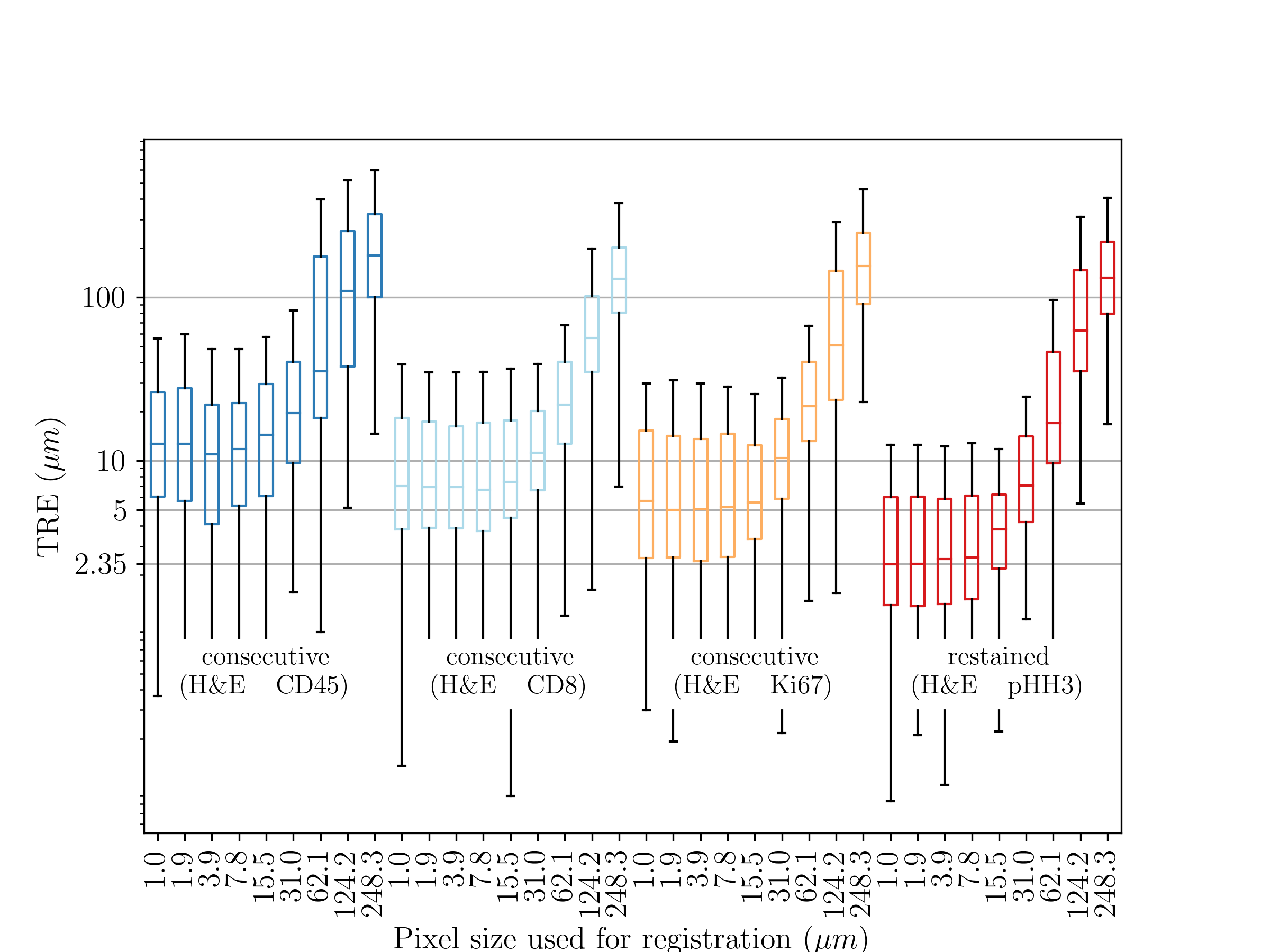}
\caption{TRE after \textbf{consecutive and re-stained} deformable
registration at different image resolutions of the HyReCo subset A for
different staining pairs (logarithmic plot). The TRE between re-stained
section (right group, red) is lower than between consecutive sections
(three left groups). The accuracy after registration of consecutive
section depends on their distance (section order: H\&E, Ki67, CD8,
CD45). The boxes denote the interquartile range and the whiskers extend
this range by a factor of 1.5.
\label{fig:deformable_hyreco_vs_restained}}
\end{figure}

\begin{figure}
\centering
\includegraphics{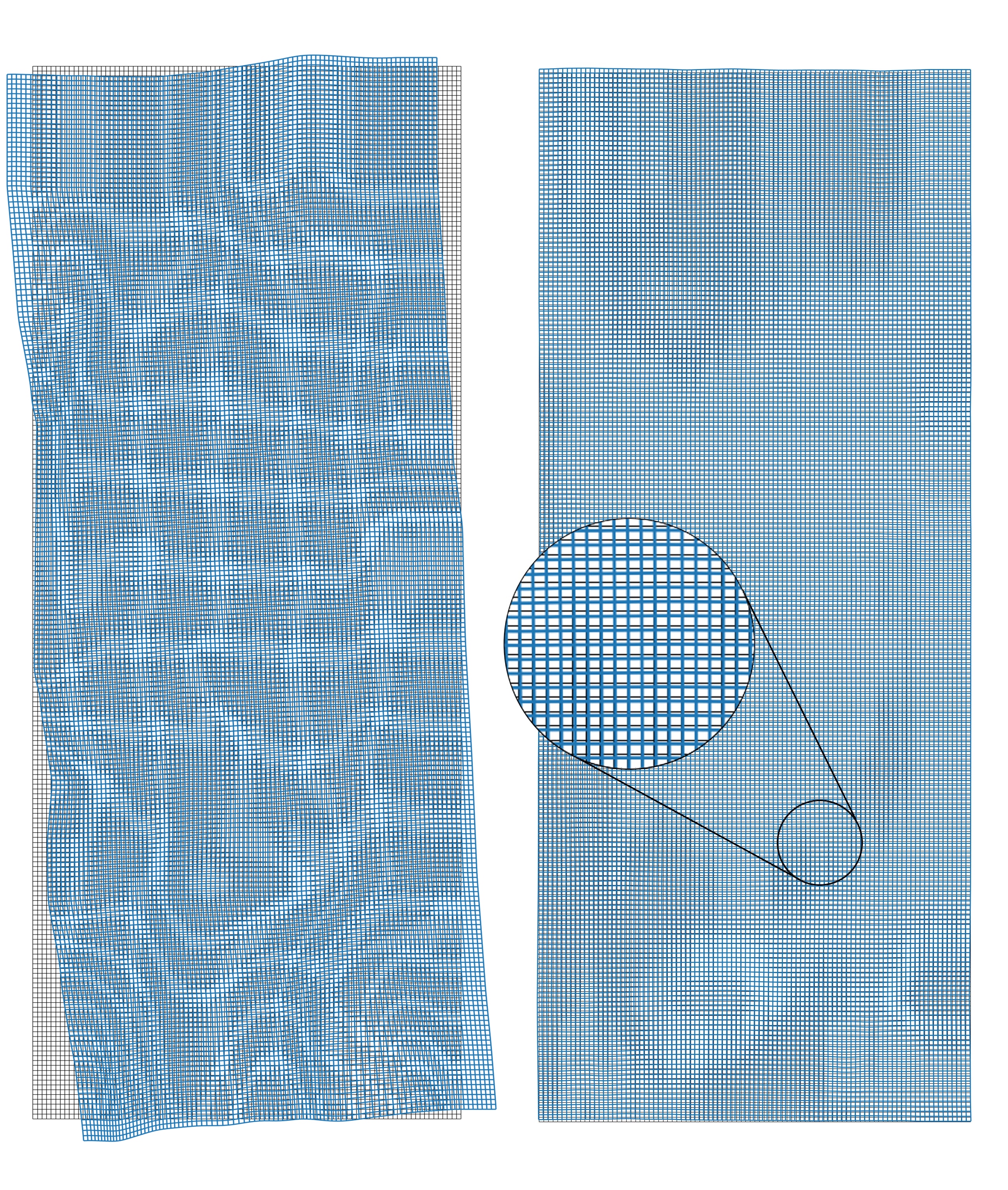}
\caption{Deformation of a consecutive (left) and a re-stained
registration (right) based on the same H\&E-stained slide (image stack
361 of the HyReCo dataset). In each image the deformation is applied to
a regular grid (background, gray) and plotted in blue (foreground). The
consecutive pair shows a larger nonlinear component but small non-linear
effects are also visible between the two re-stained images.
\label{fig:deformations}}
\end{figure}

Histological images are typically stored at different image resolutions
in a pyramidal image format to accommodate for the large size of the
images and to make the different scales of tissue structures easily
accessible. The registration can be computed at any of these scales and
the result can be interpolated to apply it to higher image resolutions.
We measure the registration accuracy with respect to the resolution used
for registration and compare affine and deformable registration on
consecutive and re-stained sections.

\hypertarget{consecutive-sections}{%
\subsubsection{Consecutive sections}\label{consecutive-sections}}

The resulting landmark errors after applying the full 3-step
registration to the consecutive HyReCo subset A and to the ANHIR dataset
are shown in Fig. \ref{fig:resultsdeformableconsecutive}.

Comparing different image resolutions, a smaller pixel size is
correlated with a smaller registration error up to a level of saturation
that differs between datasets. This saturation level is likely
influenced by the quality of the slide and the similarity of the slide
pairs. The similarity is reduced with a growing distance between two
consecutive sections and small structures can no longer be aligned if
their counterpart is not present in the other slide. At image
resolutions below 2~\(\upmu\)m/px we even observe a small increase in
TRE in some datasets. This is likely due to the larger influence of
smaller structures that---due to the differences from slide to
slide---lack a correspondence and that are otherwise invisible at
coarser image resolutions.

Comparing the HyReCo to the ANHIR cases, the overall MTRE is larger in
the ANHIR dataset where the larger average landmark errors (denoted by
\(\diamondsuit\) in Fig. \ref{fig:resultsdeformableconsecutive})
indicate a higher amount of badly aligned landmark outliers. This is
likely due to the larger structural differences between the slides in
some of the ANHIR subsets (Fig. \ref{fig:anhir_structure}).

Fig. \ref{fig:images_anhir} shows one of the ANHIR image pairs after
pre-alignment, affine registration, and deformable registration.

\hypertarget{re-stained-sections-compared-to-consecutive-sections-on-the-same-tissue-block}{%
\subsubsection{Re-stained sections compared to consecutive sections on
the same tissue
block}\label{re-stained-sections-compared-to-consecutive-sections-on-the-same-tissue-block}}

Re-stained sections show very little differences and allow a
nucleus-level alignment that can be used for a multiplexed analysis of
the finest structures in the image (Fig.
\ref{fig:checkerboard_re-stained}).

The TRE in the re-stained images in HyReCo subset A reaches
2.3~\(\upmu\)m and is approximately two to four times lower than between
consecutive sections (Fig. \ref{fig:deformable_hyreco_vs_restained},
Table \ref{tbl:TRE_deformable}). As expected, the deformation between
consecutive image pairs shows stronger non-linear components than
between re-stained sections. No foldings were detected in the
deformations in any of the re-stained image pairs. A visual comparison
of the deformations after re-stained and after consecutive registrations
is shown in Fig. \ref{fig:deformations}.

From the landmark errors in the consecutive sections we are able to
derive the likely section order (HE, Ki67, CD8, CD45RO): as the distance
between two sections in the stack grows, the landmark error increases as
well. The registration accuracy in consecutive sections largely depends
on the quality and similarity of the sections. We again observe a slight
decrease in accuracy at resolutions below~2~\(\upmu\)m/px which is only
present in the consecutive but not in the re-stained subset.

\hypertarget{experiment-2-deformable-compared-to-affine-registration}{%
\subsection{Experiment 2: Deformable Compared to Affine
Registration}\label{experiment-2-deformable-compared-to-affine-registration}}

The two images of the re-stained section pair show the same tissue
specimen before and after an additional chemical processing and
scanning. We show that deformable registration leads to superior results
despite the tissue being fixed at the glass slide during re-staining. To
this end we compare the MTRE after affine and deformable registration in
all datasets.

\hypertarget{improved-accuracy-of-deformable-registration-in-all-datasets}{%
\subsubsection{Improved accuracy of deformable registration in all
datasets}\label{improved-accuracy-of-deformable-registration-in-all-datasets}}

\begin{figure}
\centering
\includegraphics{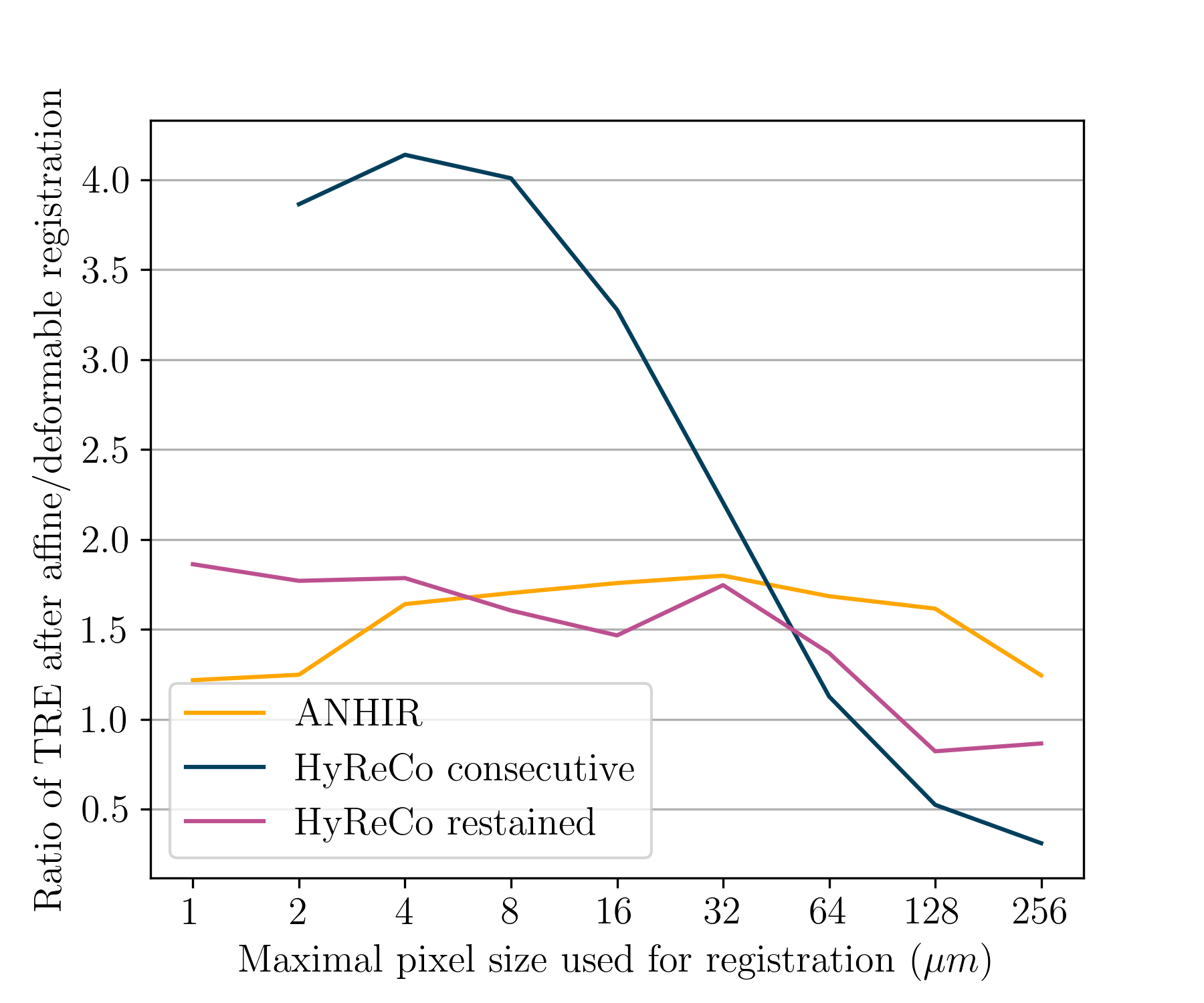}
\caption{Ratio \(\frac{\text{MTRE affine}}{\text{MTRE deformable}}\) for
ANHIR and the consecutive and re-stained HyReCo datasets. Deformable
registration outperforms affine registration except for image
resolutions coarser than 64~\(\upmu\)m/px.
\label{fig:affine_vs_deformable_ratio}}
\end{figure}

Deformable registration outperforms affine registration except for image
resolutions coarser than 64~\(\upmu\)m/px in all datasets (Fig.
\ref{fig:affine_vs_deformable_ratio}, Table
\ref{tbl:affine_vs_deformable}). Compared to consecutive sections, the
difference between affine and deformable registration is lower in the
re-stained dataset which is due to the smaller mechanical deformation in
the processing. The lower difference in the ANHIR dataset compared to
the consecutive subset of HyReCo is likely due to the larger proportion
of artifacts and structures without correspondence in this dataset.

\begin{longtable}[]{@{}
  >{\raggedright\arraybackslash}p{(\columnwidth - 4\tabcolsep) * \real{0.4364}}
  >{\raggedright\arraybackslash}p{(\columnwidth - 4\tabcolsep) * \real{0.2364}}
  >{\raggedright\arraybackslash}p{(\columnwidth - 4\tabcolsep) * \real{0.3273}}@{}}
\caption{Best median TRE obtained and required image resolution. The
MTRE between re-stained sections is lower by a factor of approx. 2. MTRE
increases with the distance between the sections for consecutive
sections. \label{tbl:TRE_deformable}}\tabularnewline
\toprule()
\begin{minipage}[b]{\linewidth}\raggedright
pair of stains
\end{minipage} & \begin{minipage}[b]{\linewidth}\raggedright
best MTRE
\end{minipage} & \begin{minipage}[b]{\linewidth}\raggedright
im. resolution
\end{minipage} \\
\midrule()
\endfirsthead
\toprule()
\begin{minipage}[b]{\linewidth}\raggedright
pair of stains
\end{minipage} & \begin{minipage}[b]{\linewidth}\raggedright
best MTRE
\end{minipage} & \begin{minipage}[b]{\linewidth}\raggedright
im. resolution
\end{minipage} \\
\midrule()
\endhead
HE--pHH3 (restained) & 2.3 \(\upmu\)m & 1.0 \(\upmu\)m/px \\
HE--Ki76 & 5.4 \(\upmu\)m & 3.9 \(\upmu\)m/px \\
HE--CD8 & 6.7 \(\upmu\)m & 7.8 \(\upmu\)m/px \\
HE--CD45RO & 11.0 \(\upmu\)m & 3.9 \(\upmu\)m/px \\
\textbf{all HyReCo consecutive} & 7.1 \(\upmu\)m & 3.9 \(\upmu\)m/px \\
ANHIR & 16.0 \(\upmu\)m & 1.0 \(\upmu\)m/px \\
\bottomrule()
\end{longtable}

\begin{longtable}[]{@{}
  >{\raggedright\arraybackslash}p{(\columnwidth - 4\tabcolsep) * \real{0.5323}}
  >{\raggedright\arraybackslash}p{(\columnwidth - 4\tabcolsep) * \real{0.2097}}
  >{\raggedright\arraybackslash}p{(\columnwidth - 4\tabcolsep) * \real{0.2581}}@{}}
\caption{Best MTRE obtained after affine and deformable registration.
\label{tbl:affine_vs_deformable}}\tabularnewline
\toprule()
\begin{minipage}[b]{\linewidth}\raggedright
pair of stains
\end{minipage} & \begin{minipage}[b]{\linewidth}\raggedright
MTRE affine
\end{minipage} & \begin{minipage}[b]{\linewidth}\raggedright
MTRE deformable
\end{minipage} \\
\midrule()
\endfirsthead
\toprule()
\begin{minipage}[b]{\linewidth}\raggedright
pair of stains
\end{minipage} & \begin{minipage}[b]{\linewidth}\raggedright
MTRE affine
\end{minipage} & \begin{minipage}[b]{\linewidth}\raggedright
MTRE deformable
\end{minipage} \\
\midrule()
\endhead
ANHIR & 19.6 \(\upmu\)m & 16.0 \(\upmu\)m \\
HyReCo consecutive (subset A) & 28.0 \(\upmu\)m & 6.8 \(\upmu\)m \\
HyReCo restained (subset B) & 1.60 \(\upmu\)m & 0.86 \(\upmu\)m \\
\bottomrule()
\end{longtable}

\hypertarget{superiority-of-deformable-registration-in-a-separate-re-stained-dataset}{%
\subsubsection{Superiority of deformable registration in a separate,
re-stained
dataset}\label{superiority-of-deformable-registration-in-a-separate-re-stained-dataset}}

In the separate subset B of re-stained slides (H\&E-PHH3) where no
consecutive sections are available, the MTRE is lower and reaches
0.86~\(\upmu\)m which is at the same level as the intra-observer error.
The difference to subset A is likely influenced by the pairwise landmark
setup.

\begin{figure}
\centering
\includegraphics{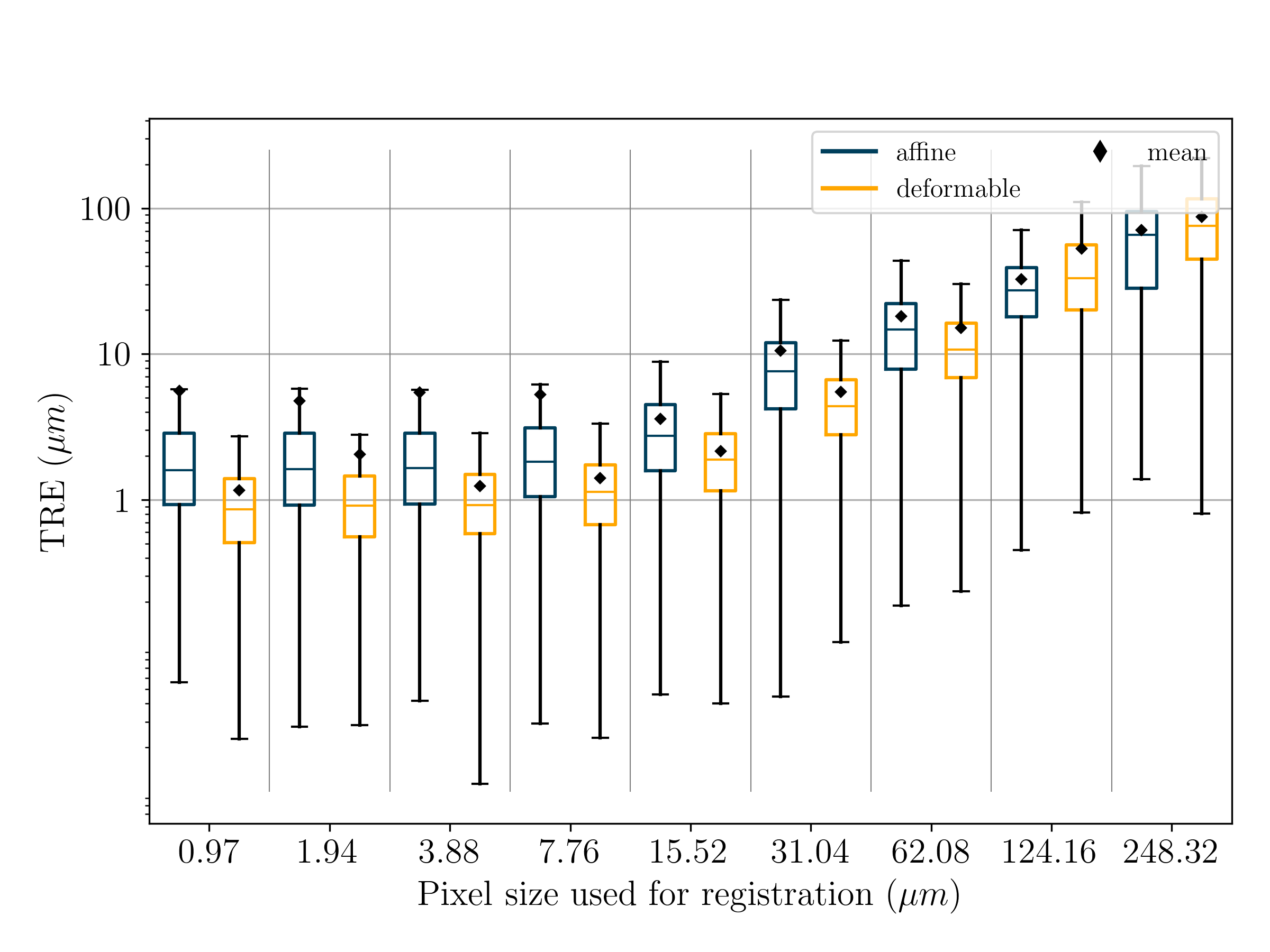}
\caption{Median TRE of \textbf{re-stained} image pairs after affine and
deformable registration at different image resolutions (HyReCo subset
B). Deformable registration does further improve the landmark error if
compared to affine registration. \label{fig:results_restained_hyreco}}
\end{figure}

In subset B, the deformable registration again lowers the landmark error
compared to affine registration (Fig.
\ref{fig:results_restained_hyreco}, but to a lower degree than between
consecutive sections (0.86 \(\upmu\)m compared to 1.60 \(\upmu\)m). A
visualization of the deformation field after re-stained section
registration shows a small non-linear component which is consistent with
the lower landmark error (Fig. \ref{fig:deformations}).

We note that purely re-stained sections are easier to annotate than
consecutive sections because the corresponding structures can easily be
identified. This leads to a lower TRE in HyReCo subset B compared to
subset A.

The better correspondence of the two sections leads to an additional
advantage of re-stained sections that cannot be measured in terms of
landmark error: Since landmarks in consecutive sections have only been
placed on corresponding structures, areas without correspondence are not
reported and therefore not part of the TRE. This is a limitation of the
current approach but could at least partly be mitigated by resorting to
a different measurement of alignment, such as the difference of
segmentations or larger structures.

\hypertarget{computation-times-of-deformable-compared-to-affine-registration}{%
\subsection{Computation Times of Deformable Compared to Affine
Registration}\label{computation-times-of-deformable-compared-to-affine-registration}}

The computation time of an image registration algorithm depends largely
on the implementation and on the size of the input images but also on
other factors like CPU and RAM performance, disk access etc. We report
the measurements of our setup (Intel(R) Core(TM) i7-7700K CPU (4.20GHz,
four cores) with 32 GB of RAM) in order to give a relative comparison
with respect to the size of the images.

For an affine registration on the HyReCo data\footnote{We do not
  systematically report the computation times in the ANHIR data because
  of a large number of different image sizes which makes the comparison
  of the computation times difficult.}, the average computation time
ranges from 0.5 seconds (image size 400 x 800, 62.1 \(\upmu\)m/pixel) to
58 seconds (image size 12800 x 25600, 1.94 \(\upmu\)m/pixel). The
majority of the computation time for large images is spent on the
deformable registration. Previous analyses
\autocite{konig_matrix-free_2018-1} have shown that doubling the image
resolution (an increase of four times the number of pixels) leads to a
four-fold increase in the computation time. In other words, the
computation time is roughly linearly dependent on the number of pixels
in the image. Together with the contributions from pre-alignment and
affine registration, we see a similar trend in the computation times in
Table \ref{tbl:computation_time}. For deformable registration on the
HyReCo data, the average computation time ranges from 2.6 seconds to 30
minutes.

\begin{longtable}[]{@{}
  >{\raggedleft\arraybackslash}p{(\columnwidth - 4\tabcolsep) * \real{0.3421}}
  >{\raggedleft\arraybackslash}p{(\columnwidth - 4\tabcolsep) * \real{0.3289}}
  >{\raggedleft\arraybackslash}p{(\columnwidth - 4\tabcolsep) * \real{0.3289}}@{}}
\caption{Execution time for a deformable registration of consecutive
sections with respect to image size (and resolution). Larger image sizes
require a larger computation time.
\label{tbl:computation_time}}\tabularnewline
\toprule()
\begin{minipage}[b]{\linewidth}\raggedleft
im. res. (\(\upmu\)m/px)
\end{minipage} & \begin{minipage}[b]{\linewidth}\raggedleft
mean exec. time (s)
\end{minipage} & \begin{minipage}[b]{\linewidth}\raggedleft
approx. image size (px)
\end{minipage} \\
\midrule()
\endfirsthead
\toprule()
\begin{minipage}[b]{\linewidth}\raggedleft
im. res. (\(\upmu\)m/px)
\end{minipage} & \begin{minipage}[b]{\linewidth}\raggedleft
mean exec. time (s)
\end{minipage} & \begin{minipage}[b]{\linewidth}\raggedleft
approx. image size (px)
\end{minipage} \\
\midrule()
\endhead
248.32 & 3.1 & 100 x 200 \\
124.16 & 2.9 & 200 x 400 \\
62.08 & 2.6 & 400 x 800 \\
31.04 & 3.7 & 800 x 1600 \\
15.52 & 8.3 & 1600 x 3200 \\
7.76 & 24 & 3200 x 6400 \\
3.88 & 89 & 6400 x 12800 \\
1.94 & 310 & 12800 x 25600 \\
0.97 & 1831 & 25600 x 51200 \\
\bottomrule()
\end{longtable}

\hypertarget{discussion}{%
\section{Discussion}\label{discussion}}

We compared the accuracy of numerical image registration in re-stained
and consecutive sections in histopathology. The median landmark error in
re-stained sections goes down to 0.86~\(\upmu\)m. When compared on the
same tissue block, the registration error between re-stained sections is
smaller by a factor of two to five compared to the corresponding
consecutive sections (2.3 \(\upmu\)m compared to 7.1 \(\upmu\)m). In
consecutive sections, the accuracy largely depends on the sections
quality and image resolution.

The difference in the alignment quality between re-stained and
consecutive sections is relevant for applications where small structures
or single nuclei are of interest. An MTRE of 1.0 \(\upmu\)m allows
nucleus-level alignment which is infeasible in serial sections where the
same nucleus is often not present on the next slide. For comparison, the
size of an average mammalian nucleus is approx. 6
\(\upmu\)m~\autocite{alberts_molecular_2002}, while tissue sections
typically measure 2--5 \(\upmu\)m in thickness. The increased accuracy
comes at the price of the loss of the physical stained glass slide and
an increased processing time due to the de-staining. Only the staining
that is applied last can be conserved physically. Especially in clinical
settings, long-term storage of the glass slides and short time to
diagnosis are important.

Smaller nonlinear deformations occur in the re-staining process, likely
due to the mechanical and chemical manipulation and tile stitching
during scanning. These nonlinear components can also be observed in the
deformation fields resulting from the registration of re-stained images.
When aiming at a high registration accuracy in re-stained images,
deformable image registration further decreases the landmark error in
fine image resolutions.

The accuracy of the registration depends on the employed image
resolution. As histological images are typically organized in a
pyramidal structure, lower-resolution representations can be extracted
without additional computational effort. Otherwise, loading the image
into memory in order to produce a low-resolution representation further
extends the computation time.

In our experiments, the impact of the image resolution was highest in
re-stained sections where optimal results could be reached at 0.97
\(\upmu\)m/px. Finer image resolutions even exceeded 180 GB of RAM on a
more powerful computer. In consecutive sections, the gain in accuracy of
the registration stagnates between 7.8 \(\upmu\)m/px and 3.89
\(\upmu\)m/px such that these registrations can be computed based on
smaller image size and hence require less memory and time. We assume
that fine structures that lack correspondence have a negative impact on
registration accuracy in finer image resolutions.

Our analysis is limited by the focus on landmarks as the only
measurement of accuracy. Since the landmarks were placed at positions
that can be re-identified by a human observer, these locations likely
have a superior contrast and thus have a higher impact on the distance
measure. This could lead to a bias in the evaluation that underestimates
the registration error in low-contrast regions.

The purely landmark-based approach also ignores the quality of the
alignment of larger structures. This could be included by segmenting
corresponding areas in multiple slides and evaluating the alignment of
these segmentations.

The regularity or smoothness of the deformation is another quality
criterion for an image registration. We automatically analyze the
deformed grid for folds (one occurrence in 36 cases for HyReCo subset A,
zero occurrences in subset B) but otherwise did not systematically
evaluate smoothness of the deformation except for visual inspection.

\hypertarget{conclusion}{%
\section{Conclusion}\label{conclusion}}

In conclusion, re-stained sections allow an accurate registration of
differently stained structures that is below the level required to align
single nuclei. Registrations of consecutive sections result in a higher
alignment error that increases with the distance between the slides.
Consecutive sections are better suited to align larger areas such as
tumor or inflammatory areas based on a second stain. We recommend
deformable registration which was always more accurate, and the use of
re-stained sections, if possible. Higher image resolutions benefit the
accuracy, as long as the increase in image detail leads to an increase
in corresponding structures.

\hypertarget{disclosures}{%
\section{Disclosures}\label{disclosures}}

Jeroen van der Laak was a member of the advisory boards of Philips, the
Netherlands and ContextVision, Sweden, and received research funding
from Philips, the Netherlands, ContextVision, Sweden, and Sectra, Sweden
in the last five years. He is chief scientific officer (CSO) and
shareholder of Aiosyn BV, the Netherlands.

The authors have no relevant financial interests in the manuscript and
no other potential conflicts of interest to disclose.


\printbibliography

\newpage

{
\centering
\thispagestyle{empty}

Visual Abstract \\ \ \\

\includegraphics[]{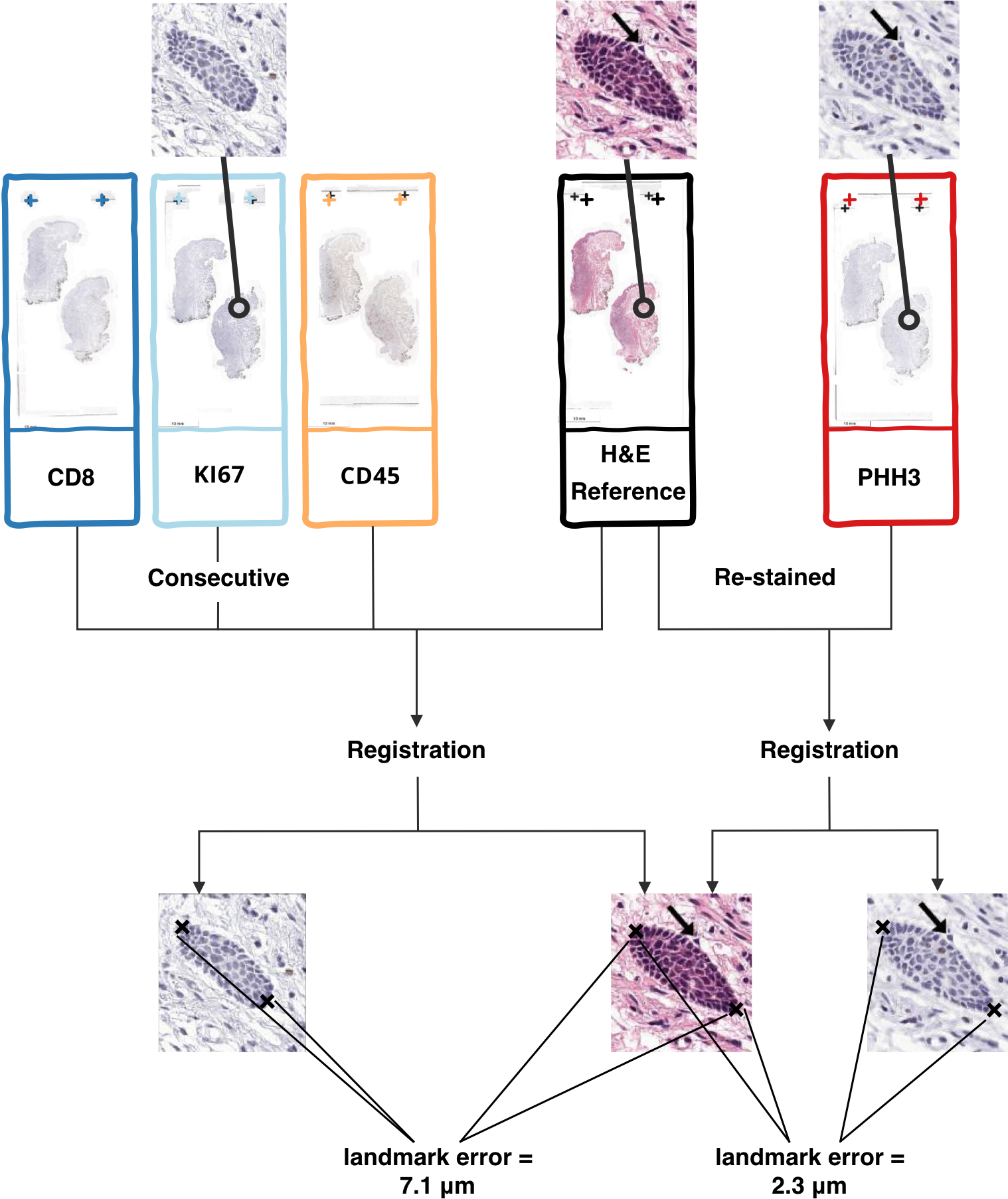}

}

\end{document}